\documentclass{article}
\usepackage{amsmath}
\usepackage{amsfonts}
\begin{document}

\title{Spin in a General Time Varying Magnetic Field: \\Generalization of the Adiabatic Factorization of Time Evolution}

\author{J. Chee\\\\Department of Physics, \\Tianjin Polytechnic University,
 Tianjin 300387, China}

 \date{October, 2010}
 \maketitle
 \begin{abstract}
 An extension of the adiabatic factorization of the time evolution
 operator is studied for spin in a general time varying magnetic
 field ${\bf B}(t)$. When ${\bf B}(t)$ changes adiabatically,
 such a factorization reduces to the product of the geometric operator which embodies the Berry phase phenomenon
 and a usual dynamical operator. For a general time variation of ${\bf B}(t)$, there should be another operator $N(t)$ in the factorization 
 that is related to non-adiabatic transitions. A simple and explicit expression
 for the instantaneous angular velocity of this operator is derived. This is done in a way that
 is independent of any specific representation of spin.
 Two classes of simple conditions are given under which the
 operator $N(t)$ can be made explicit. As a special case, a generalization of the traditional magnetic resonance condition is pointed out.
  
 \end{abstract}
 \pagebreak

\section{Introduction}
The purpose of this paper is to study a unique factorization of the
time evolution operator for spin in a general time varying magnetic
field that should be seen as a generalization of the adiabatic
factorization.

Let us first recall that in the proof of the quantum adiabatic
theorem as presented in standard texts such as Messiah
\cite{messiah}, a path-dependent operator $A(t)$ and a dynamical
operator $D(t)$ are successively extracted from the time evolution
operator. Then, there is left an operator $N(t)$, the
product of which with $A(t)$ and $D(t)$ should equal the time
evolution operator $U(t)$, i.e., $U(t)=A(t)D(t)N(t)$. The rest of
the proof of the quantum adiabatic theorem amounts to the proof of
the statement that $N(t)$ approaches the identity operator in the
adiabatic limit. The nature of the path-dependent operator has been
clarified with the concept of the geometric phase \cite{berry,
simon, shapere}. Near the adiabatic limit, the operator $N(t)$ (we shall sometimes refer to 
it as the non-adiabatic operator)
can be analyzed in the spirit of perturbative analysis \cite{berry1,
moody, hagedorn}. For a general time variation of the parameters in
the Hamiltonian, the non-adiabatic operator $N(t)$ in the time evolution
operator is responsible for transitions among instantaneous
eigenstates. While the operator $A(t)$ is characterized by the
geometry of the path traversed by the set of parameters in the
Hamiltonian, and $D(t)$ is the usual dynamical operator, a simple
characterization of the operator $N(t)$ that corresponds to a
general time variation of the parameters in the Hamiltonian seems to
be still lacking.

This paper intends to clarify the structure of the operator $N(t)$
for spin in a general time varying magnetic field from a perspective
that is independent of perturbation theory. The question we ask is
how the geometry of the path of the direction of the magnetic field
and the dynamical effect contribute in determining $N(t)$. As it
turns out, $N(t)$ in this case is determined by an angle parameter
associated with the geometry of the
path, a dynamical angle that characterizes $D(t)$ and the speed
function that describes how the path is traversed in time. While
this is reminiscent of the fact that the operator $N(t)$, which is
an $SU(2)$ rotation, should in principle be characterized by three
parameters, our point in this paper is that the instantaneous
angular velocity of $N(t)$ can be explicitly expressed in a simple
way by using the time dependent parameters associated with the
Hamiltonian. The factorization of the time evolution operator $U(t)$
derived here can be applied to the most general time variation of
the magnetic field, including the adiabatic scenario as a special
case. Through this natural factorization of $U(t)$, we find that
there exist new explicit solutions of time evolution that go beyond
the scope of perturbative analysis.

In a certain sense, the approach we adopt in this paper differs from
the traditional adiabatic perturbation theory (See, for example,
\cite{moody} for a review). Let us observe that in the proof of the
adiabatic theorem \cite{messiah}, the path-dependent operator is
formally constructed out of the energy eigenstates, i.e.,
\begin{equation}
A(t)=\sum_m|\psi_m(t)\rangle\langle \psi_m(0)|,
\end{equation}
where $|\psi_m(t)\rangle$ is the instantaneous eigenstate of $H(t)$
that satisfies
\begin{equation}
\langle \psi_m(t)|\dot \psi_m(t)\rangle=0.
\end{equation}
Such a formal expression in terms of eigenstates is necessary
because in the general situation no specific information of the
Hamiltonian is available. The traditional adiabatic perturbation
theory is strongly influenced by the use of instantaneous
eigenststes. While such an approach is quite general in nature and
have wide applications, it can also lead to complicated calculations
and the results obtained from such an approach cannot be
extrapolated to situations that are not within the scope of
perturbation theory. We are of the point of view that in analyzing a
specific problem such as the spin in a time varying magnetic field,
the use of energy eigenstates can be avoided and that the time
evolution can be studied by exploring the algebraic structure of the
Hamiltonian alone in constructing a factorization of the time
evolution that can then be applied to any representation.

The Hamiltonian for a spin ${\bf S}$ with magnetic moment $-k{\bf
S}$ in a time varying magnetic field is
\begin{equation}
H=k{\bf B}(t)\cdot {\bf S}=kB(t){\bf n}(t)\cdot{\bf S},
\end{equation}
where $k$ is a constant, ${\bf B}(t)=B(t){\bf n}(t)$ is the magnetic
field and ${\bf n}(t)$ is a unit vector.

If ${\bf B}(t)$ changes adiabatically, the time
evolution operator takes the form
\begin{equation}
U_A(t)=A(t)D(t),
\end{equation}
where
\begin{equation}
A(t)= T\exp \big(-i\int_{0}^{t}({\bf n}\times \dot{\bf n})\cdot{\bf
S}~d\tau\big),
\end{equation}
and
\begin{equation}
D(t)=\exp\big(-i\int_{0}^{t}kB(\tau){\bf n}(0)\cdot {\bf
S}~d\tau\big).
\end{equation}
The operator $A(t)$ is determined by the path traversed by the
direction of the magnetic field ${\bf n}(t)$ on the unit
two-dimensional sphere. (We set $\hbar=1$.)

For the spin 1/2 ~case, the operator $A(t)$ is written in this way
using Pauli matrices in Berry's iterative approach to calculate
non-adiabatic corrections to the geometric phase \cite{berry1}. In
the factorization of time evolution studied in the present paper,
the two factors $A(t)$ and $D(t)$, and another factor $N(t)$ to be
derived later belong to an arbitrary finite-dimensional irreducible
representation of $SU(2)$, so the method and result here apply to
any spin.

First, let us fix certain notations. Recall from the quantum
mechanics of spin that the dot product in the Hamiltonian indicates
an isomorphism between the three dimensional vector space and the
Lie algebra with a certain basis: $\{-iS_1, -iS_2, -iS_3\}$. To be
more specific, $\{-iS_1, -iS_2, -iS_3\}$ here is a basis of an
arbitrary finite dimensional irreducible representation of the Lie
algebra $su(2)$. So, in effect, the operators that are concerned in
this paper can be thought of as finite dimensional square matrices.

Let ${\bf e}_1(t)$, ${\bf e}_2(t)$, and ${\bf e}_3(t)$ be three orthogonal unit
vectors determined by the differential equations
\begin{equation}
{\dot {\bf e}}_i(t)=~({\bf n}\times \dot{\bf n})\times {\bf e}_i(t),
\end{equation}
with $i=1,2,3$, and the initial condition
\begin{equation}
{\bf e}_1(0)={\bf n'}(0),~~ {\bf e}_2(0)={\bf n}(0)\times{\bf n'}(0),~~{\bf e}_3(0)={\bf n}(0),
\end{equation}
where the prime in ${\bf n'}(0)$ means derivative taken with respect to arc
length. The vectors ${\bf e}_1(t)$ and ${\bf e}_2(t)$ satisfy the condition
$\dot{\bf e}_1(t)\cdot{\bf e}_2(t)=0$. When viewed as tangent vectors, they have the geometric
meaning of being parallel transported along the curve ${\bf n}(t)$ on the unit two-sphere.

Let us define $S_1,~S_2$ and $S_3$ as follows:
\begin{equation}
S_1={\bf e}_1(0)\cdot{\bf S},~~S_2={\bf e}_2(0)\cdot {\bf S},~~ S_3={\bf e}_3(0)\cdot {\bf S}.
\end{equation}
In accordance with the quantum mechanics for spin, these operators are assumed to
satisfy the usual commutation relations $[S_k, S_l]=~i\epsilon_{klm}S_m$.

Finally, the operator $A(t)$ is, by definition, the solution to
the differential equation
\begin{equation}
{\dot A}(t)=~(-i({\bf n}\times \dot{\bf n})\cdot {\bf S})A(t),
\end{equation}
with the initial condition that $A(0)$ is equal to the identity.
Because $\dot A^{-1}A + A^{-1}\dot A =0$, we have $\dot A^{-1}=~-A^{-1}{\dot A}A^{-1}$. Then it follows from the
definition of $A(t)$ that
\begin{equation}
{\dot A}^{-1}(t)=A^{-1}(t)(i({\bf n}\times \dot{\bf n})\cdot {\bf S}).
\end{equation}

\section{Mathematical Framework}
In this section, we give an introduction to existing knowledge on
the fundamental properties of the operator $A(t)$ and also a
rotation angle between two frames naturally associated with the path
traversed by the magnetic field. This preparation clarifies the
general factorization of time evolution completed in the next
section.

\subsection{Two properties of $A(t)$}
It is well known \cite{simon} that the Berry phase has mathematical
interpretations in terms of fundamental concepts in differential
geometry and Lie groups. Here we review the relevant mathematics
from the point of view of the operator $A(t)$. As we shall see,
$A(t)$ plays an important role in deriving the general factorization
of $U(t)$. The two characterizing properties of $A(t)$ are as
follows.

Property 1. First, we have the following property of $A(t)$:
\begin{equation}
A^{-1}(t){\bf e}_i(t)\cdot {\bf S}A(t)~=~{\bf e}_i(0)\cdot {\bf S}=S_i.
\end{equation}
To prove this, note that this clearly holds for $t=0$ by definition of $A(t)$. Now it suffices to
prove that the left hand side is constant, or, its time derivative is zero. This can be verified
by using the expressions for $\dot A(t)$ and ${\dot A}^{-1}(t)$, the equation
${\dot {\bf e}}_i(t)=~({\bf n}\times \dot{\bf n})\times {\bf e}_i(t)$, and the formula $[{\bf a\cdot S}, {\bf{b\cdot S}}]=~i\bf{(a\times b)\cdot S}$.

This property of $A(t)$ \cite{chee1} is best understood in terms of the adjoint
representation of the Lie group $G$ generated (through the exponential map) by the Lie algebra with
the basis $\{-iS_1,-iS_2,-iS_3\}$. So $G$ is
isomorphic to $SU(2)$ or $SO(3)$. In the case where $G$ is isomorphic to $SU(2)$, it is well known that
this adjoint representation is a 2 to 1 covering map from $G$ to $SO(3)$, and so in general the holonomy associated with
the standard metric on the two sphere is insufficient in determining the Berry phase information contained in $A(t)$.
So the following property is indispensable if one wishes to cast the
solid angle Berry phase from the point of view of $A(t)$.

Property 2. For a cyclic change of ${\bf n}(t)$ with
${\bf n}(T)={\bf n}(0)$, we have $A(T)=\exp[-i\Omega{\bf
n}(0)\cdot{\bf S}]$ , where $\Omega$ is the solid angle subtended by
the closed path at the origin of the parameter space. Here the parameter is $\bf B$.

This is equivalent to associating a
solid angle Berry phase to each individual energy eigenstate.
To prove this property, let $|\psi_m({\bf n}(0))\rangle$ be an initial eigenstate of ${\bf n}(0)\cdot {\bf S}$, i.e.,
\begin{equation}
{\bf n}(0)\cdot {\bf S}|\psi_m({\bf n}(0))\rangle= m|\psi_m({\bf n}(0))\rangle.
\end{equation}
Then it can be shown, using Property 1, that
$|\psi_m({\bf n}(t))\rangle=A(t)|\psi_m({\bf n}(0))\rangle$
satisfies
\begin{equation}
{\bf n}(t)\cdot {\bf S}|\psi_m({\bf n}(t))\rangle=~m|\psi_m({\bf n}(t))\rangle, ~~~\langle \psi_m({\bf n}(t))|\dot \psi_m({\bf n}(t))\rangle=0.
\end{equation}
Now one may choose a locally
single valued $|m({\bf n})\rangle$ such that $|\psi_m({\bf n}(t))\rangle=e^{i\gamma (t)}|m({\bf n}(t))\rangle$.
Then $e^{i\gamma (t)}$ for a closed path can be calculated to be $e^{-im\Omega}$ by Berry's phase two-form method. Since
the $|\psi_m({\bf n}(0))\rangle$'s form a complete basis for the representation space, it is clear that $A(T)=\exp[-i\Omega{\bf
n}(0)\cdot{\bf S}]$.

In conclusion, although the operator $A(t)$ has been written only as a time ordered exponential,
it nevertheless should be seen as an operator that can be explicitly constructed for a given representation of $SU(2)$ and
for a given ${\bf n}(t)$.
This can be seen as a consequence of the fact that the instantaneous angular velocity has the special form
$({\bf n}\times \dot{\bf n})$, so that the above two properties hold for $A(t)$.
In a specific representation, one has $A(t)=\sum_m|\psi_m(t)\rangle\langle \psi_m(0)|$. Because the matrices $S_1$,$S_2$, and $S_3$
can now be written explicitly, one can obtain explicit solutions to the eigenvalue equation
${\bf n}(t)\cdot {\bf S}|\psi_m({\bf n}(t))\rangle=~m|\psi_m({\bf n}(t))\rangle$.
The condition $\langle \psi_m({\bf n}(t))|\dot \psi_m({\bf n}(t))\rangle=0$ then fixes a phase factor for the
eigenfuction and $A(t)$ can then be made explicit. This contrasts the non-adiabatic operator $N(t)$ to be discussed later:
$N(t)$ can be made explicit only in special cases, although the instantaneous angular velocity of $N(t)$ has an explicit expression.

\subsection{Relative rotation between two triads}

In using an iterative method to calculate non-adiabatic corrections to the spin 1/2 geometric phase,
Berry considered the relative rotation between the two triads $({\bf e}_1(t), {\bf e}_2(t), {\bf e}_3(t))$ and
$({\bf n'}(t),{\bf n}(t)\times{\bf n'}(t), {\bf e}_3(t))$ \cite{berry1}. The relation between the two triads can be expressed as
\begin{equation}
{\bf e}_1(t)=~{\bf n'}(t)\cos\beta+{\bf n}(t)\times{\bf n'}(t)\sin\beta ,
\end{equation}
\begin{equation}
{\bf e}_2(t)=~-{\bf n'}(t)\sin\beta+{\bf n}(t)\times{\bf n'}(t)\cos\beta .
\end{equation}
The condition ${\bf e'}_1\cdot{\bf e}_2=0$ then implies that
\begin{equation}
{\beta'}(t)=-({\bf n}(t)\times{\bf n'}(t))\cdot{\bf n''}(t).
\end{equation}
So $\beta (t)$ can be written as
\begin{equation}
\beta(t)=\int_0^{l(t)}-{\bf n''}\cdot({\bf n}\times{\bf n'})ds,
\end{equation}
where $l(t)$ is the arc length traveled through by ${\bf n}(t)$.
Thus, $\beta(t)$ is a geometric quantity in  the sense that it is
determined by the path traversed by ${\bf n}(t)$ on the two-sphere and is independent of how the path is parameterized.

If $l(t)$ increases with $t$, then the relation between ${\bf
n'}(t)$ and the time derivative $\dot{\bf n}(t)$ is $\dot{\bf
n}(t)={\bf n'}(t)|\dot{\bf n}(t)|$. Then we also have
\begin{equation}
\beta(t)=\int_0^{t}-\frac{\ddot{\bf n}}{\dot{|\bf n|}^2}\cdot({\bf n}\times\dot{\bf n})dt.
\end{equation}
So $\beta$ is explicitly determined if ${\bf n}(t)$ is given.

\section{General factorization of time evolution}
In this section we study how the adiabatic factorization can be
generalized when the magnetic field varies in a general way. The purpose
is to derive an explicit expression of the angular velocity of the non-adiabatic operator
in terms of the relative rotation $\beta(t)$, which is a geometric quantity, and the dynamical quantity
$\phi(t)=-\int_{0}^{t}kB(\tau)~d\tau$ in a way that is independent of specific
representations of spin.

Let the non-adiabatic operator be $N(t)$ so that the time evolution
operator in the general situation is
\begin{equation}
U(t)=A(t)D(t)N(t).
\end{equation}
Substituting this expression into the Schrodinger equation
\begin{equation}
i{\dot U}(t)=H(t)U(t)
\end{equation}
we obtain
\begin{equation}
iA^{-1}\dot{A}DN+i\dot{D}N+iD\dot{N}=A^{-1}HADN.
\end{equation}
By the first property of the operator $A(t)$ from last section, it is clear that
$i\dot{D}N=A^{-1}HADN$, so that we have
\begin{equation}
A^{-1}\dot{A}DN+D\dot{N}=0,
\end{equation}
or,
\begin{equation}
\dot{N}=~-D^{-1}A^{-1}\dot{A}DN=~iD^{-1}A^{-1}({\bf n}\times \dot{\bf n})\cdot{\bf
S}ADN.
\end{equation}
This equation determines the operator $N(t)$ that satisfies the initial condition $N(0)=I$.

To calculate $iD^{-1}A^{-1}({\bf n}\times \dot{\bf n})\cdot{\bf
S}AD$ in the above expression, let
\begin{equation}
{\bf n}(t)\times \dot{\bf n}(t)=a(t){\bf e}_1(t)+~b(t){\bf e}_2(t).
\end{equation}
Then the relative orientation of the two triads from last section tells us that
\begin{equation}
a(t)=|\dot{\bf n}(t)|\sin\beta(t),~~ b(t)=|\dot{\bf n}(t)|\cos\beta(t).
\end{equation}
By Property 1 for $A(t)$, we have
\begin{equation}
A^{-1}\dot{A}=~-i(a(t)S_1+~b(t)S_2).
\end{equation}
Using the formulas
\begin{equation}
e^{-i\phi S_3}S_1e^{i\phi S_3}=S_1\cos\phi+S_2\sin\phi,
\end{equation}
\begin{equation}
e^{-i\phi S_3}S_2e^{i\phi S_3}=S_1(-\sin\phi)+S_2\cos\phi,
\end{equation}
we now have
\begin{equation}
-D^{-1}A^{-1}\dot{A}D=ia(t)(S_1\cos\phi+S_2\sin\phi)+ib(t)(S_1(-\sin\phi)+S_2\cos\phi).
\end{equation}
where
\begin{equation}
\phi(t)=-\int_{0}^{t}kB(\tau)~d\tau.
\end{equation}
The expression for $-D^{-1}A^{-1}\dot{A}D$ can now be written in the form
\begin{equation}
-D^{-1}A^{-1}\dot{A}D=iS_1|\dot{\bf n}(t)|\sin[\beta(t)-\phi(t)]+iS_2|\dot{\bf n}(t)|\cos[\beta(t)-\phi(t)],
\end{equation}
from which we get
\begin{equation}
N(t)=T\exp\Big(-i\int_0^t\big(\omega_1(\tau)S_1+\omega_2(\tau)S_2\big)d\tau\Big),
\end{equation}
where the two non-vanishing components of the instantaneous angular velocity are
\begin{equation}
{\omega_1}(t)=-|\dot{\bf n}(t)|\sin[\beta(t)-\phi(t)],~ ~{\omega_2}(t)=-|\dot{\bf n}(t)|\cos[\beta(t)-\phi(t)].
\end{equation}
The expression for $N(t)$ in terms of the explicitly determined
angular velocity is the main result of this paper. $|\dot{\bf n}(t)|$ here has the
simple meaning of being the velocity at which the path $\bf n$ on the two-sphere is traversed in time.
Together with $|\dot{\bf n}(t)|$, the two naturally defined quantities $\beta(t)$
and $\phi(t)$, one geometrical and the other dynamical, provide a
simple expression for the angular velocity which in turn completely
determines the operator $N(t)$.

Since $S_1$ and $S_2$ can be expressed in terms of the raising and lowering operators $S_{+}=S_1+iS_2$, and $S_{-}=S_1-iS_2$, $N(t)$
also can be written as
\begin{equation}
N(t)=T\exp\Big(\int_0^t\big(z(\tau)S_{+}-z^{*}(\tau)S_{-}\big)d\tau\Big),
\end{equation}
where
\begin{equation}
z(t)=\frac{|\dot{\bf n}(t)|}{2}e^{i[\beta(t)-\phi(t)]}.
\end{equation}
In a certain sense, $N(t)$ here is analogous to the non-adiabatic operator that we derived for the
Landau problem with a general time dependent electric field, where it is a time-ordered
coherent state displacement operator that describes inter-Landau level transitions \cite{chee}. However, the time-ordered coherent state
displacement operator there can be explicitly determined because the commutation relation $[a,  a^\dagger]=1$ is relatively simple.
In fact, $N(t)$ for this time varying Landau problem can be shown to be equal to a numerical phase factor times the ordinary exponential.
In contrast, the time-ordered exponential here has no immediate relation to the ordinary exponential. This is due to the fact that we
have a different commutation relation between $S_{+}$ and $S_{-}$, i.e., $[S_{+}, S_{-}]=2S_3$. We shall demonstrate below that $N(t)$
can be made explicit for special cases where additional conditions are specified.

\section{Two classes of explicit factorizations}
The general factorization includes previous studied cases such as
the adiabatic approximation and the sudden approximation as special
cases. It may also be applied to study in detail certain
generalizations of the quantum adiabatic theorem such as when the
parameter passes through the degeneracy point \cite{avron}.

Let us take the sudden approximation for example. It states that if
the parameters in the Hamiltonian changes (even if appreciably) in a
sufficiently short time, then under quite general conditions, the
quantum state remains close to the initial state \cite{messiah}.
Suppose now that the magnetic field keeps its magnitude $B$ constant
but changes direction in a very short time, then the dynamical
quantity $\phi(t)=-\int_{0}^{t}kB(\tau)~d\tau$ is close to zero
(because $t\rightarrow 0$) and thus can be ignored. We thus have
$D(t)=1$. If $\phi(t)=0$, we have
\begin{eqnarray}
N(t)&=&T\exp\Big(-i\int_0^t\big(\omega_1(\tau)S_1+\omega_2(\tau)S_2\big)d\tau\Big),\\
    &=&T\exp\Big(i\int_0^t|\dot{\bf n}(\tau)|\sin\beta(\tau)S_1+|\dot{\bf
    n}(t)|\cos\beta(\tau)(\tau)S_2\big)d\tau\Big),\\
    &=&T\exp\Big(i\int_0^ta(\tau)S_1+b(\tau)S_2\big)d\tau\Big).
\end{eqnarray}
We thus have
\begin{eqnarray}
\dot{N}(t)&=&(a(t)S_1+b(t)S_2)N(t),\\
          &=&(({\bf n}(t)\times \dot{\bf n}(t))\cdot{\bf e}_1(t)S_1+({\bf n}(t)\times \dot{\bf n}(t))\cdot{\bf
          e}_2(t)S_2)N(t).
\end{eqnarray}
On the other hand, by using Property 1. in section 2, we have
\begin{eqnarray}
{\dot A}^{-1}(t)&=&A^{-1}(t)(i({\bf n}\times \dot{\bf n})\cdot {\bf
S}),\\
                &=&(A^{-1}(t)(i({\bf n}\times \dot{\bf n})\cdot {\bf
S})A(t))A^{-1}(t),\\
&=&(({\bf n}(t)\times \dot{\bf n}(t))\cdot{\bf e}_1(t)S_1+({\bf
n}(t)\times \dot{\bf n}(t))\cdot{\bf
          e}_2(t)S_2)A^{-1}(t).
\end{eqnarray}
Since $N(t)$ and $A^{-1}(t)$ satisfy the same differential equation
with the same initial condition, they must be the same:
$N(t)=A^{-1}(t)$. From this we deduce that $U(t)=1$. The
factorization therefore recovers the sudden approximation as a
special case.

The explicit form of the angular velocity of $N(t)$ also suggests
that $N(t)$ itself can be made explicit under certain conditions
that go beyond previously studied situations. Below we list two
simple classes of explicit solutions of $N(t)$.

(i) The operator $N(t)$ is explicit if $\beta(t)-\phi(t)\equiv0$. This condition can be
written more explicitly, by the definitions
of $\beta(t)$ and $\phi(t)$, as the following:
\begin{equation}
kB(t)=-\frac{\ddot{\bf n}(t)}{|\dot{\bf n}(t)|^2}\cdot({\bf n}(t)\times\dot{\bf n}(t)).
\end{equation}
Under this condition, it is clear that $N(t)$ becomes
an ordinary exponential so that it is explicitly determined:
\begin{equation}
N(t)=\exp\Big(i\int_0^t|\dot{\bf n}(\tau)|S_2d\tau\Big)=\exp(iS_2l(t)),
\end{equation}
where $l(t)$ is the length of the path traversed by ${\bf n}(t)$ on the unit sphere.
The condition can be seen as putting no constraint on how ${\bf
n}(t)$ varies while giving $B(t)$ explicitly for a given function
${\bf n}(t)$. Therefore, there are infinitely many explicit examples
that belong to this class. For instance, if ${\bf
n}(t)=(\sin\theta\cos\omega t, \sin\theta\sin\omega t, \cos\theta)$,
then this condition automatically gives $kB=-\omega\cos\theta$.

This class is clearly outside the scope of perturbative analysis. We
recall that near the adiabatic limit, the dynamical angle changes much more rapidly
than $\beta(t)$ does so that the oscillating sine and cosine
functions can make sure, under quite general conditions, that the small effect of $|\dot{\bf
n}(t)|\sim\epsilon$ does not accumulate during an adiabatic process
with $t\in[0, 1/\epsilon]$. The expression for $N(t)$ in this class
of examples goes beyond adiabatic perturbation theory in that even
if $|\dot{\bf n}(t)|$ is small, it may accumulate over time (because
there are no oscillating factors in the intergrand) so that
transitions among instantaneous eigenstates can be large. In fact, we see that in this case
the transition probability among instantaneous eigenstates is characterized by the length of the path
traversed by ${\bf n}(t)$ at time $t$.

In general, ${\bf n}(t)$ can be an arbitrary smooth curve on the
2-sphere and it is also possible that $B(t)$ changes sign during the
time evolution. This only agrees with our assumption that the
magnetic field is allowed to change in a general way which includes
the situation where the magnetic field ${\bf B}(t)=B(t){\bf n}(t)$
passes through the degeneracy point of the Hamiltonian. In order
that $B(t)$ and ${\bf n}(t)$ change smoothly, $B(t)$ is allowed to
become zero and to change sign across the degeneracy point.

(ii) The second class of explicit solutions for $N(t)$ is obtained under the conditions that
\begin{equation}
|\dot{\bf n}(t)|=c_1, \and~~~ \beta(t)-\phi(t)=c_2t,
\end{equation}
where $c_1$ and $c_2$ can be arbitrary constants. If these two
conditions are satisfied, we have
\begin{equation}
\dot N=ic_1(S_1\sin c_2t+S_2\cos c_2t)N=ic_1e^{ic_2tS_3}S_2e^{-ic_2tS_3}N.
\end{equation}
Let
\begin{equation}
N=e^{ic_2tS_3}N_1.
\end{equation}
Then we have
\begin{equation}
{\dot N}_1=(ic_1S_2-ic_2S_3)N_1.
\end{equation}
From this we see that $N_1$ is explicitly solved and an explicit expression for $N(t)$ is obtained:
\begin{equation}
N(t)=e^{ic_2tS_3}e^{(ic_1S_2-ic_2S_3)t}.
\end{equation}

This class of explicit factorizations include as a special case the
well-known exact solution developed specifically for analyzing the
phenomenon of magnetic resonance. In this case the function ${\bf
n}(t)$ is simply ${\bf n}(t)=(\sin\theta\cos\omega t,
\sin\theta\sin\omega t, \cos\theta)$ and so we have
$kB=-\omega\cos\theta-c_2$, which is an arbitrary constant. The
traditional method for solving this special case is different from
the factorization of time evolution we studied here. This is hardly
surprising because this special solution was found long before the
discovery of the geometric phase. Let us note that this special case
also belongs to the first class if we choose $c_2=0$. Here, one must
carefully distinguish between the condition $kB=-\omega\cos\theta$
and the traditional magnetic resonance condition. In the traditional
magnetic resonance analysis, one considers transitions among energy
levels along the direction $(0,0,1)$ fixed by a static magnetic
field. The energy levels correspond to the static magnetic field
only, not the total Hamiltonian, which also contains an oscillating
magnetic field. In our notation, this static magnetic field is then
$-kB\cos\theta$. (Let us assume $k<0$ here.) The traditional
magnetic resonance condition is  $\omega=-kB\cos\theta$. The
difference lies in that the condition $\omega=-kB\cos\theta$ is the
resonance condition in the fixed direction while the condition
$kB=-\omega\cos\theta$ here represents an analogous resonance in the
moving direction represented by ${\bf n}(t)$ and the energy levels
along ${\bf n}(t)$ are determined by the whole instantaneous
Hamiltonian. These two resonance conditions only agree with each
other for small $\theta$. When $\theta$ is close to $\pi/2$, the
dominating effect would be transitions along the moving direction
${\bf n}(t)$. It may be interesting to study if the condition
$kB=-\omega\cos\theta$ can have experimental consequences.

We want to point out that there are new examples that belong to this
class. Now we shall give a new  explicit example of ${\bf n}(t)$ and
$B(t)$ that satisfy Equation (47). First we note that the condition
$\beta(t)-\phi(t)=c_2t$ amounts to
$kB(t)={\dot\beta}(t)-c_2=-\frac{\ddot{\bf n}(t)}{|\dot{\bf
n}(t)|^2}\cdot({\bf n}(t)\times\dot{\bf n}(t))-c_2$, so that we can
focus on finding ${\bf n}(t)$ that satisfies $|\dot{\bf n}(t)|=c_1$.
Let us take ${\bf n}(t)=(\sin\lambda t\cos\varphi(t), \sin\lambda
t\sin\varphi(t), \cos\lambda t)$. Then it is easily shown that
\begin{equation}
|\dot{\bf n}(t)|^2={\lambda}^2+{\dot\varphi}^2{\sin^2\lambda t}=c_{1}^2.
\end{equation}
For $\lambda>0$, $c_{1}^2>{\lambda}^2$ and $0\leq t<\frac{\pi}{2\lambda}$, the solution for $\varphi(t)$ that satisfies $\varphi(0)=0$
can be found to be
\begin{equation}
\varphi(t)=\pm\frac{1}{\lambda}\sqrt{c_{1}^2-\lambda^2}\ln\big(\tan(\frac{\lambda t}{2}+\frac{\pi}{4})\big).
\end{equation}
From these solutions, $kB(t)=-\frac{\ddot{\bf n}(t)}{|\dot{\bf
n}(t)|^2}\cdot({\bf n}(t)\times\dot{\bf n}(t))-c_2$ is also
explicitly known. In this example, it is clear that the curve ${\bf
n}(t)$ in general does not lie in a plane.

It could be of interest to find more examples within or beyond these
two simple classes listed above and to investigate possible
experimental manifestations of the solutions obtained.

\section{Concluding Remarks}
The main purpose of this paper is to clarify $N(t)$ for a much
studied physical model: spin in a general time varying magnetic
field. Our motivation is to understand, in this specific example,
the interplay between the geometric effect and the dynamical effect
(as represented by the operators $A(t)$ and $D(t)$ respectively) in
determining the structure of the $N(t)$. Our main
result is the explicit construction of the instantaneous angular
velocity of $N(t)$. This clarifies the structural information of
$N(t)$ and facilitates the study of transition probabilities among
instantaneous eigensates for a general time evolution. It puts
$N(t)$ on an equal footing with the operator $A(t)$ and $D(t)$ in
the sense that the instantaneous angular velocities of all these
operators are now explicitly known. With this structural information
of $N(t)$, we show that $N(t)$ itself can be made explicit under two
classes of simple conditions, both of them go beyond the scope of
perturbation theory. This gives new physical examples of explicit
functions of the magnetic field ${\bf B}(t)$ that correspond to the
explicit construction of the time evolution operator.

The study of the extension of the adiabatic factorization of $U(t)$ in the above spirit can be useful in other physical situations too.
In a recent work, we examined the extension of the adiabatic factorization of the time evolution for the Landau problem with a
spatially uniform electric field ${\bf E}(t)$ that has a general time
dependence, including the situation where ${\bf E}(t)$ changes direction in time \cite{chee}.
The factorization in this
case shows that all the three factors can be determined explicitly:
The non-adiabatic factor summarizes all
non-adiabatic effects in the form of a path-ordered coherent state
displacement operator while the geometric operator is a path-ordered
magnetic translation. The path-ordered operators imply the
existence of two numerical phase factors that are quantum mechanical
in nature. And a simple expression is obtained for non-adiabatic
transitions among different energy levels (inter Landau level
transitions) for a general time evolution using the non-adiabatic
operator. This time dependent Landau problem has been analyzed before from the point of view of constructing
the quantum propagator \cite{dodonov}. Such a procedure of constructing the propagator however, does not seem to point to a factorization
of the time evolution operator that can make the time evolution transparent.

It could be of interest to study the factorization of time evolution
for other physical systems, classical or quantum mechanical, where
the same line of ideas applied here may contribute to the
elucidation of the physical and/or mathematical aspects involved.


\begin{thebibliography}{07}
 \bibitem{messiah} A. Messiah, \textsl{Quantum Mechanics} (North Holland, Amsterdam, 1970), Vol. 2.
 \bibitem{berry} M.V. Berry,  Proc. R. Soc. A392, 45-57 (1984).
 \bibitem{simon} B. Simon,  Phys. Rev. Lett. 51, 2167-2170 (1983).
 \bibitem{shapere} \textsl{Geometric Phases in Physics}, edited by A Shapere and F Wilczek, World Scientific (1989).
 \bibitem{berry1}  M.V. Berry, Proc. R. Soc. A414, 31-46 (1987).
 \bibitem{moody} J. Moody, A. Shapere and F. Wilczek, Adiabatic Effective Lagrangians, in \textsl{Geometric Phases in Physics}, World Scientific (1989).
 \bibitem{hagedorn} G.A. Hagedorn and A. Joye,  Communications in Mathematical Physics, Volume 250, Issue 2, pp.393-413 (2004).
 \bibitem{chee1} J. Chee, Phys. Letts. A 275,  473-480 (2000).
 \bibitem{chee} J. Chee, Annals of Physics, 324, 97-105 (2009).
 \bibitem{avron} J. E. Avron and A. Elgart, Communications in Mathematical Physics 203 (2): 445¨C463 (1999).
 \bibitem{rabi} I.I. Rabi, N. F. Ramsey and J. Schwinger, Rev. Mod. Phys. 26, 167-171 (1954).
 \bibitem{dodonov} V.V. Dodonov, I.A. Malkin and V.I. Man'ko, Physica, 59, 241-256 (1972).
\end{thebibliography}
\end{document}